\documentclass[reqno]{birkmult}
\usepackage{latexsym}
\usepackage{amssymb}
\usepackage{amsmath, graphicx, amscd}

\theoremstyle{plain}

\theoremstyle{definition}

\numberwithin{equation}{section}
\numberwithin{theorem}{section}

\def\half{\frac{1}{2}} 
\def\lap{\bigtriangleup}            
\def\empty{\phi}                    

\def\R{\mathbb R}            %
\def\D{\mathbb D}            %
\def\C{\mathbb C}        %
\def\H{\mathbb H}            %

\begin{document}

\title[Duplex numbers, diffusion systems, and generalized quantum mechanics]{Duplex numbers, diffusion systems,\\
                       and generalized quantum mechanics}

\author{Jerzy Kocik}
\address{{\bf Current:} Department of Mathematics, Southern Illinois University, Carbondale, IL62901\\
                {\bf At the publication time:} Department of Physics, University of Illinois at Urbana-Champaign, Urbana, IL61801}
\email{jkocik@siu.edu}

\thanks{Preprint of an article published in the {\it Int. J. Theo. Phys.}, Vol. 38 No. 8, 1999 (pp. 2221-30).
Although the text is not altered, the author now prefers term ``hyperbolic numbers" over the original "duplex numbers". }

\begin{abstract} 
We show that the relation between the Schr\"odinger equation
and diffusion processes has an algebraic nature and can be revealed via
the structure of ``duplex numbers."  This helps one to clarify that
quantum mechanics cannot be reduced to diffusion theory.
Also, a generalized version of quantum mechanics where $\C$ is replaced
by a normed algebra with a unit is proposed.
\end{abstract}


\maketitle

\section*{Introduction}  \label{s:1}

The idea that
extending the formalism of quantum mechanics beyond the field of
complex numbers may bring about some additional interesting structure
has been investigated since the 1960s
(Finkelstein {\it et al.}, 1962; Emch, 1963; Nash and Joshi, 1992;
Adler, 1995;	and references therein).
The argument that the extended field must be a division algebra
limits the investigations via the Frobenius (1878) theorem	to three fields:
real numbers $\R$, the complex plane $\C$, and quaternions $\H$.
One can however argue that the object replacing complex numbers may be
any associative algebra with a quadratic norm, not necessarily positive
definite.  This leads to a wide spectrum of interesting
possibilities, among which the Clifford algebras may serve as the
simplest generalization
(encompassing quaternions and duplex numbers investigated below).

\smallskip

Nagasawa (1993) introduced a certain type of diffusion system and showed
that it can replace the Schr\"odinger equation under appropriate change of
potential.  This observation led him to conclude that quantum
mechanics can be demystified and replaced by a more ``intuitive" and
less mysterious diffusion theory.

In this paper, we introduce ``quantum mechanics" over duplex numbers,
i.e., with the imaginary unit $I$ satisfying $I^2=+1$.
We show that the Nagasawa type of diffusion systems
may be understood within the quantum formalism over the field of duplex
numbers.
In this framework, we can see why, contrary to some claims,
the standard quantum mechanics cannot be reduced to a diffusion
system, and the argument is topological in nature.
[This is relevant to the question of whether ``one needs
$i$ in quantum mechanics" (Jauch, 1973; Jammer, 1974).]

The example presented here shows also that the formalism of quantum
mechanics may be extended beyond the paradigm of division fields $\R$,
$\C$, and $\H$.  Some comments on generalized quantum mechanics
follow.

\section{The algebra of duplex numbers}  \label{s:1}


Duplex numbers were introduced by Clifford (1873, 1878) as ``double
numbers" and they recently gained some interest among physicists
[see, e.g., Hucks (1993) for the relation to Dirac spinors,
and Kunstatter {\it et al.} (1983) for applications in the
theory of gravitation].

By duplex numbers we understand the algebra
$\D = \{\; a+bI\;\mid\;a,b\in\R\;\}$ with $I^2=+1$.
In many respects, $\D$ is similar to the field of
complex numbers $\C$, except that the elements of the form $a\pm aI$ do not
have an inverse. The conjugation of a duplex number $z=a+bI$ is defined
as $\bar z =a-bI$, and a norm is defined via
$$
\mid z\mid^2=z\bar z=(a+bI)(a-bI)=a^2-b^2
\eqno(1.1)
$$
Thus, duplex numbers $\D$ form a plane $\R^2={\rm span}\,\{1,I\}$ with
hyperbolic geometry.  The inverse of $z$ is $z^{-1}=\bar z/\mid z\mid^2$.
Duplex numbers can be expressed in the polar form
$$
           z=\rho\;e^{I\varphi} =
           \rho (\cosh \varphi + I\cdot \sinh\varphi)
\eqno(1.2)
$$
with $\tanh\varphi=b/a$ and $\rho\in\R$ for numbers of positive square
norm, $\mid z\mid^2>0$, and $\rho\in I\cdot\R$ otherwise.
In particular, a hyperbolic version of de Moivre's formula holds
$$
  \rho_1 e^{I\varphi_1} \;\cdot\; \rho_2 e^{I\varphi_2} =
  \rho_1 \rho_2 e^{I(\varphi_1 + \varphi_2)}
\eqno(1.3)
$$
Duplex numbers manifest two-dimensional space-time structure in an
algebraic form.
In particular, elements of form
$$
  e^{I\varphi}
\eqno(1.4)
$$
represent hyperbolic rotations of the plane $\D$ (``boosts") and
represent the connected component of the Lie group $SO(1,1)$,
containing the group unit.

%

\section{Quantum mechanics over duplex numbers}  \label{s:2}

The standard Schr\"odinger equation of quantum mechanics over complex
numbers is
$$
         i\partial_t\Psi + {1\over 2} \lap \Psi - U\cdot \Psi = 0
\eqno(2.1)
$$
($m=1$, $\hbar=1$) where $\Psi:\,\R^n\to\C$ is a complex-valued wave
function, the Laplacian $\lap=\nabla\cdot\nabla$ is a composition of
gradients, and $U:\,\R^n\to\R$ is a potential.
We may express the wave function in polar (logarithmic) coordinates as
$\Psi=e^{R+iS}$.	It is a well-known result (Nelson 1966; Pelce, 1996)
that the Schr\"odinger equation resolves into a pair of partial differential
equations for the real-valued functions $R$ and $S$:
$$
\begin{cases}
  -\partial_t S + \half\lap R +\half(\nabla R)^2 - \half (\nabla S)^2 - U = 0 \\[5pt]
   \partial_t R + \half\lap S + \nabla S \cdot\nabla R  = 0              \cr
\end{cases}
\eqno(2.2)
$$
Indeed, let $\Psi=e^z$.  Then (2.1) is equivalent to
$i\partial_t z + \half\lap z +\half(\nabla z)^2 -  U = 0 $.
Now, substituting $z=R+iS$ and separating the real and the imaginary
parts gives the two equations of (2.2).
\\

Consider now an analog of the Schr\"odinger equation over duplex numbers,
$$
     I\partial_t\empty + {1\over 2} \lap \empty - W\cdot \empty = 0
\eqno(2.3)
$$
for some potential $W:\R^n\to\R$.  Assume $\empty=e^{R+IS}$.  Then the
duplex Schr\"odinger equation resolves into a pair of diffusion-type
equations
$$
\begin{cases}
   \partial_t S + \half\lap R + \half(\nabla R)^2+\half (\nabla S)^2 - W = 0 \\[5pt]
   \partial_t R + \half\lap S + \nabla S \cdot\nabla R  = 0           
\end{cases}
\eqno(2.4)
$$
This result may be restated in a notation that encompasses both the
complex and the duplex case.
A quantum wave function $\Psi:\;\R^n\to \R^2$ assumes values in
two-dimensional algebra with a unit, spanned by $\{1,a\}$.
\\

\noindent
{\bf Theorem 1:} {\sl
Define a (generalized) Schr\"odinger equation
$$
        {\mathcal S}[a;U]\; \Psi[a;R,S]=0
\eqno(2.5a)
$$
where ${\mathcal S}$ is an operator based on an (invertible)
algebraic unit $a$,
$$
   {\mathcal S}[a;U] = a^{-1}\partial_t + {1\over 2} a^{-2}\lap  + U\cdot
\eqno(2.5b)
$$
and the generalized wave function has polar form
$$
        \Psi[a;R,S]=e^{R+aS}
\eqno(2.5c)
$$
Then if $a^2=-1$,  ${\mathcal S}[a,U]$ is the standard Schr\"odinger
operator and (2.5a) describes a quantum system with potential $U$.
If $a^2=1$, then ${\mathcal S}[a,W]$ is the ``duplex Schr\"odinger operator"
describing the diffusion system (2.4).
}
\\

The question is whether these two systems can be equivalent.  Comparison
of (2.2) and (2.4) leads immediately to the following.

\bigskip\noindent
{\bf Proposition 2:} {\sl
The solutions to these two equations
$$
\begin{array}{rl}
      (i)  &\quad {\mathcal S}[i,U]\; \Psi[i,R,S] = 0 \qquad \ \ i^2=-1  \\[5pt]
      (ii) &\quad {\mathcal S}[I,W]\; \Psi[I,R',S'] = 0 \qquad I^2=+1  
\end{array}
\eqno(2.6)
$$
coincide, i.e., $S\equiv S'$ and $R\equiv R'$, if
$$
              W=U + 2\partial_tS+(\nabla S)^2      \eqno(2.7a)
$$
or, equivalently, if
$$
             W=-U + \lap R + (\nabla R)^2    \eqno(2.7b)
$$
}

The equivalence of Schr\"odinger's equation (2.1) and the diffusion system
(2.4) upon condition (2.7a) is studied by Nagasawa (1993).  Here, we have
established that this relation is algebraic and originates in the
opposition of the complex versus duplex numbers.

Here is an algebraic difference between $\D$- and $\C$-quantum
mechanics.
\\

\noindent
{\bf Corollary 3:} {\sl
The duplex Schr\"odinger equation (2.4) can be written in an isotropic form
$$
\begin{cases}
   \partial_t(S+R) + \half\lap(S+R) + \half(\nabla (S+R))^2 = W \\[5pt]
   \partial_t(S-R) + \half\lap(S-R) + \half(\nabla (S-R))^2 = 0 
\end{cases}
\eqno(2.8a)
$$
which, by introducing vectors ${\bf Z} = [S+R, S-R]$ and ${\bf W}=[W,0]$,
we can be express as
$$
     \partial_t{\bf Z} + \half\lap{\bf Z} + \half(\nabla {\bf Z})^2 = {\bf W}
\eqno(2.8b)
$$
}

This can easily be verified directly from (2.4).  Notice that the
standard (complex) Schr\"odinger system does not admit such a
representation.
The symmetry of (2.8) is due to the geometry of duplex numbers; indeed,
two isotropic elements:
$$
  \gamma={\scriptstyle 1\over\sqrt{2}}\; (1+I)
  \qquad\qquad
  \bar\gamma= {\scriptstyle 1\over\sqrt{2}}(1-I)
$$
satisfy the following multiplication table
$$
              \gamma^2=\gamma
       \qquad \bar\gamma^2=\bar\gamma
       \qquad \gamma\cdot\bar\gamma=0
\ .
$$
Units $\gamma$ and $\bar\gamma$ determine the isotropic coordinates
(``light cone") on $\D$  (algebraically, each generates an ideal in $\D$).
For any element $z=\alpha\gamma + \beta\bar\gamma$, one has
$$
  \bar z = \beta\gamma + \alpha\bar\gamma
$$
with the norm
$$
  z\bar z =   (\alpha\gamma + \beta\bar\gamma)
            \;(\beta \gamma + \alpha\bar\gamma)
          = \alpha\beta(\gamma + \bar\gamma)
          = \sqrt{2}\alpha\beta
\ .
$$
Now, we can see that Equation (2.8) corresponds to the split
$R+IS={1\over\sqrt{2}}(R+I)\gamma+{1\over\sqrt{2}}(R-I)\bar\gamma$.

It is easy to see that the above two cases exhaust all two-dimensional
cases.  Let the `imaginary' element $a$ be of a general form, such that
$a^2=\alpha + a\beta = [\alpha, \beta]$.  Setting $\alpha\not= 0$ assures
invertibility of $a$.
Any specification of the vector $[\alpha,\beta]$ is equivalent to
one of two cases
$$
\begin{array}{rl}
  [-1,0] \qquad&{\mathop{\longrightarrow}}\quad\hbox{standard QM} \quad(a=i) \\[4pt]
  [+1,0] \qquad&{\mathop{\longrightarrow}}\quad\hbox{diffusion process} \quad(a=I) 
\end{array}
\eqno(2.9)
$$
discussed above (2.2, 2.4).
Indeed, transformation of basis $\{1,{\mathbf a}\}$ into a new basis
$\{1,{\mathbf f}\}$ with $\mathbf f$ defined as
$$
{\mathbf f}=\begin{cases}
            \frac{2\mathbf a -\beta}{\sqrt{|\beta^2+4\alpha|}}      & \quad\hbox{if}\quad \alpha\not=-\beta^2/2 \\[5pt]
                              {\mathbf a }-\beta/2                                           &\quad\hbox{otherwise}              
                    \end{cases}
$$
establishes an algebra isomorphism to one of the cases (2.9), since
${\mathbf f}^2= 1\cdot{\rm sgn}\,(\beta^2+4\alpha)$.

Direct calculations show that in the general basis (2.5) separates into
a system of two equations:
$$
\begin{array}{rl}
  (i)\quad& \partial_t R + \half\lap S + \nabla S \cdot\nabla R
                    +\beta(\partial_t S + \half (\nabla S)^2+U) = 0\\[5pt]
 (ii)\quad& \half\lap R +\half(\nabla R)^2
                     +\alpha(\partial_t S + \half (\nabla S)^2  = 0
\end{array}
\eqno(2.10)
$$
For the sake of illustration, consider the idempotent case $[0,1]$, i.e.,
$a^2=a$:
$$
\begin{array}{l}
   \lap R + (\nabla R)^2 = 2W                      \\[5pt]
   \partial_t (R+S) + \half\lap S + \half(\nabla S)^2
                            + \nabla S\nabla R =0
\end{array}
\ .
\eqno(2.11)
$$
The equivalence of (2.11) with the standard Schr\"odinger system
(in the sense of Proposition~2) can be ensured by $W\equiv U$ and
$2\partial_tS+ (\nabla S)^2=0$.
This case is isomorphic to the diffusion case $[1,0]$,
as a substitution $b=2a-1$ shows, since $b^2=4a^2-4a+1=1$.

\goodbreak
\bigskip\bigskip\noindent
{\bf 3. Does quantum physics need $\sqrt{-1}$ ?}

\bigskip\noindent
It has been argued (Nagasawa, 1993) that the quantum formalism can be reduced
to a study of diffusion processes and that the equivalence of Schr\"odinger's
equation to diffusion systems ``demystifies" quantum mechanics
(see also Collins, 1992).
Let us look at this proposition. Since the diffusion systems in consideration
can be viewed in terms of duplex algebra, we can now restate
the question: ``Can quantum mechanics be rewritten in terms of duplex
numbers?"  That is to say, are quantum formalisms with imaginary units
$a^2=\pm1$ equivalent in their ability to describe concrete physical
systems?

Many features of the ``duplex Schr\"odinger mechanics" invite one to advocate
such a view; let us review some of them.  First, the probability density in
the probabilistic interpretation can be obtained via an equivalent of
the familiar Born formula
$$
       \Psi\bar\Psi = e^{(R+IS)} e^{(R-IS)} = e^{2R}
\eqno(3.1)
$$
which is analogous to the standard complex version.
The additivity of phases also holds, due to (1.2).
Note also that the second equations of both versions of Schr\"odinger's
equation, the complex (2.2) and the duplex (2.4), coincide;
both represent the ``continuity equation," which can be rewritten as
$$
          \partial_t P + \nabla {\bf J} = 0
\eqno(3.2)
$$
where $P$ and $\bf J$ are the scalar-valued ``density" and vector-valued
``current," respectively.  In the duplex case, they are defined
as real functions (in position and time):
$$
      P=\Psi\bar\Psi=e^{2R}  \qquad\qquad
      {\bf J}={1\over 2I}
        (\bar\Psi\nabla\Psi-\Psi\nabla\bar\Psi)=e^{2R}\nabla S
\eqno(3.3)
$$
Thus, the ``kinematic" components of the two Schr\"odinger formalisms,
complex and duplex, coincide.  The other (first) equations in (2.2) and
(2.4) are extended versions of the Jacobi-Hamilton equation of classical
mechanics, and differ in the two cases, complex and duplex.
The question is whether a particular dynamical system can be expressed
equally well by either of them.
The free particle satisfies the equivalence condition (2.7) trivially,
$W=U=0$, and
$$
      \Psi \sim e^{I(px-Et)}
\eqno(3.4)
$$
with $E=p^2/2$. Similarly, one can replace the complex (quantum) form by the
duplex (diffusion) form for any stationary system. Indeed, assume that
$\Psi$ is separable
$$
      \Psi(t,{\bf x}) = e^{-IEt}\cdot u({\bf x})
\eqno(3.5)
$$
Here we have $\partial_tS=-E={\rm const}\,$ and $\partial_tR=0$, hence the
second (continuity) equation of (2.2) is satisfied automatically.
The duplex Schr\"odinger equation (2.4) reduces to its time-independent
version
$$
     \half\nabla u({\bf x}) =  (E + W({\bf x}))\cdot u({\bf x})
\eqno(3.6)
$$
Compare it with the standard Schr\"odinger equation
$$
     \half\nabla u({\bf x}) =  (U({\bf x}) - E)\cdot u({\bf x})
\eqno(3.7)
$$
These two differ only in interpretation of the energy; an
identification $W=U-2E$ makes the two descriptions, complex and
duplex, equivalent.  For instance, $W=x^2/2$ leads to the Hermite
polynomials, as $U=-x^2/2$ does in the standard quantum mechanics.

\smallskip

These simple cases may indeed suggest that the two descriptions are
interchangeable.
However, the equivalence breaks down once one goes beyond the configuration
spaces of trivial topology, and the reason lies in
(i) the superposition principle and (ii) the different topology of the unit
circles in $\C$ and $\D$, or the different ``symmetry groups":

$$
\begin{array}{rl}
     SO(2) &= \{e^{i\varphi}\}\sim S^1 \subset \C
                            \qquad\hbox{(compact)}\\[5pt]
   SO_{\uparrow}(1,1) &= \{e^{I\varphi}\}\sim \R \subset \D
                            \qquad\hbox{(non-compact)}
\end{array}
\eqno(3.8) 
$$

\smallskip\noindent
The essence of the formalism of quantum mechanics lies in its ability to
deal with systems where a number of states can coexist in
superposition and interfere at the time of observation.
Consider the classic Young two-slit experiment as an example.
Standard cursory estimations
(with zero potential) of the intensity at position $x$ on the screen
near its center ($x=0$) leads to a sinusoidal pattern
$$
  |e^{i\varphi} +e^{i(\varphi+\delta)}|^2 \sim \cos^2(\delta/2)
\eqno(3.9)
$$
where the phase difference from the slits is approximately
$$
\delta =x\cdot d/L
$$
with $d$ the distance between slits, and $L$ the distance to the screen
$(d<\!\!<L)$.  A similar estimation (with zero potential) for the duplex
wave gives
$$
  |e^{I\varphi} +e^{I(\varphi+\delta)}|^2 \sim \cosh^2(\delta/2)
\eqno(3.10)
$$
which has one maximum at the center and vanishes as $|x|$ increases.

The fringe pattern in the two-slit experiment arises from
the phase periodicity and therefore cannot be explained by diffusion
equations.  (In the complex case, phase $S$ develops modulo $2\pi$ on
the circle, while in the duplex case, it may develop in an unbounded
manner.)
For this reason, results of any path-split type of experiment (involving
configuration space with a nontrivial fundamental group) cannot be
explained within the framework of duplex quantum mechanics.

\smallskip

In conclusion, the interpretational
problems of quantum mechanics cannot be resolved in terms of diffusion
processes (beyond simple cases of topologically trivial configuration
spaces).
These two phenomena, quantum processes and
diffusion processes, are different in nature as they correspond to
different nonisomorphic algebras $\C$ and $\D$ with topologically
different symmetry groups.

\bigskip

The above discussion is relevant to the old question of whether ``quantum
mechanics requires the imaginary unit $i$"
[see summaries of discussion in Jauch (1973) and Jammer (1974)].
It was first posed by P. Ehrenfest and (unsatisfactorily) addressed
by Pauli (1933).
Later, Stueckelberg {\it et al.} (1960; Stueckelberg and Guening, 1961)
considered real Hilbert spaces and showed that the uncertainty principle
requires a superselection rule that is equivalent to a complex
structure in the Hilbert space.
The present paper points to the topological nature of the problem and
to the compactness of $S^1\subset\C$ as the source of the experimental
results (``fringes") in the topologically nontrivial configuration spaces.

\goodbreak
\bigskip\bigskip\noindent
{\bf 4.  Quantum mechanics generalized}

\bigskip\noindent
A possible generalization of the formalism of quantum mechanics has been
advocated for some time, but has typically been limited to the division
algebras $\R$, $\C$ and $\H$ (Adler, 1995).
Starting from
the classical dynamical interpretation of the Schr\"odinger equation
(Strochi, 1966; Rowe {\it et al.}, 1980; Heslot, 1985; Jones, 1992),
Millard (1997) has recently proposed a generalization to the case of an
associative ring with a conjugation.
The present paper also encourages one to go beyond this paradigm and consider
quantum formalism based on algebras with a norm relaxed from
the condition of positive definiteness.
Here we present a general outline; a more detailed exposition will be
developed elsewhere.

Let $A$ be an associative algebra with a unit and a (quadratic) norm
$||\; ||^2$ that is not necessarily positive  definite (and possibly
degenerate).	We shall demand that
$$
|| a b ||^2 = ||a||^2\cdot ||b||^2
\eqno(4.1)
$$
which is dictated by the correspondence principle (Adler, 1995).
In an $A$-quantum mechanics, the Hilbert space is replaced by a
right $A$-module with a set of $A$-valued operators acting on the left.
In particular, the canonical commutation rules
$$
[\hat p, \hat x] = a \cdot \hbar \cdot {\rm id}
\eqno(4.2)
$$
for some fixed invertible $a\in A$ (hereafter $\hbar=m=1$) are consistent
with the Schr\"odinger representation
$$
\hat x = x\cdot
\qquad
\hat p = a^{-1}\hbar\nabla
\qquad
\hat H = -a^{-1}\hbar\partial_t
\eqno(4.3)
$$
For instance, a particle in potential $U$ is described by
$$
   -a^{-1}\partial_t \Psi ={\scriptstyle{1\over 2}}a^{-2} \lap \Psi + U\Psi
\eqno(4.4)
$$

\bigskip

Clifford algebras (Porteous, 1981) present an example of normed algebras with a unit and
therefore determine a family of generalized forms of quantum mechanics.
Let $V=\R^{(p,q)}$ be an $n$-dimensional space  equipped with a
pseudo-Euclidean structure of signature $(p,q)$ where $n=p+q$.
The corresponding Clifford algebra $\R_{(p,q)}$ can be viewed as a
$2^n$-dimensional Grassmann space $A=\wedge V$ with the algebra product
induced from
$$
  vw = - g(v,w) + v\wedge w
\eqno(4.5)
$$
for $v,w\in V\subset A$  [in particular, $vv=-g(v,v)$].
Let $\{e_i\}$ be an orthonormal basis in $V$.  Denote the ordered set of
its indices by $I=\{1,2,\ldots,n\}$.
Any index subset $A\subset I$ defines a basis element $e_A$ of $A$; for
instance
$$
         e_\empty = 1
\qquad   e_{\{i\}}=e_i
\qquad   e_{\{1,3,7\}} = e_1 e_3 e_7
\qquad   e_I = e_1 e_2 \ldots e_n
$$
where $\empty$ denotes the empty subset.

Now, let us consider Clifford quantum mechanics.
It seems natural (but not necessary) to chose $a$ of Equation (4.3) to
be the volume element (pseudoscalar) $a=e_I$  (clearly, $e_I^2=\pm1$).
Thus the commutation rules (4.2) read $[\hat p,\hat x] = e_I  {\rm id}$,
and the Schr\"odinger representation $\hat x = x\cdot$,
$\hat p = e_I^{-1}\nabla$ gives, after multiplying both sides by $e_I^2$,
$$
   - e_I\partial_t \Psi = {\scriptstyle{1\over 2}} \lap \Psi \pm U\Psi
\eqno(4.6)
$$
where the undetermined sign is that of $e_I^2=(-)^{n(n-1)/2 + p}$.
Due to the noncommutativity of Clifford algebras, the polar form of the
Schr\"odinger equation does not emerge naturally.
The Schr\"odinger equation resolves into $2^n$ intertwined equations
of a general form
$$
\partial_t \Psi_A =\pm {\scriptstyle{1\over 2}} \lap \Psi_{A^c} \pm U \Psi_{A^c}
$$
labeled by subsets $A\subset I$, where $A^c=I-A$.
Intuitively, subspaces with $e_A^2=1$ correspond to ``diffusion sectors,"
and those with $e_A^2=-1$ correspond to ``quantum sectors."

\bigskip

The two algebras $\C$ and $\D$ juxtaposed in this paper correspond to two
cases of Clifford algebras based on one-dimensional spaces (with $e_1$
identified with $i$ or $I$, respectively).
The low-dimensional cases encompass the following:
$$
\begin{matrix}
\R_{1,0} \;\longrightarrow\; \hbox{standard quantum mechanics}     \hfill&\\[2pt]
\R_{0,1} \;\longrightarrow\; \hbox{diffusion system}               \hfill&\\[2pt]
\R_{0,0} \;\longrightarrow\; \hbox{heat equation  } (a=1)          \hfill&\\[2pt]
\R_{2,0} \;\longrightarrow\; \hbox{quaternionic quantum mechanics} \hfill&
\end{matrix}
$$
In particular, quaternions correspond to the Clifford algebra of
$\R^{(2,0)}$ and have a basis
$i=e_1$,   $j=e_2$, $k=e_1\wedge e_2$.
One can express the wave function as a sum of two ``complex"-valued
functions
$$
           \Psi =\psi + k \phi
                =(\psi_1 + i\psi_2 ) + k ( \phi_1 + i\phi_2 )
\eqno(4.7)
$$
with $\psi_1$, $\psi_2$, $\phi_1$, and $\phi_2$ real.  Then, assuming
$U$ real or complex, the generalized Schr\"odinger equation resolves into
a pair
$$
\begin{array}{rl}
 \partial_t \psi &=- {\scriptstyle{1\over 2}}\lap\phi + U\phi\\[5pt]
-\partial_t \phi &=- {\scriptstyle{1\over 2}}\lap\psi + U\psi\cr
\end{array}
\ .
\eqno(4.8)
$$

\bigskip

A detailed description of Clifford quantum mechanics involves
commutation rules of Clifford algebras, which goes beyond the scope of this
paper and will be presented elsewhere.

\goodbreak
\bigskip\bigskip\noindent
{\bf Bibliography}
\bigskip

\smallskip\noindent
Adler, S.L. (1995).
{\it Quaternionic Quantum Mechanics and Quantum Fields},
Oxford University Press, Oxford.

\smallskip\noindent
Clifford, W.K. (1873).
{\it Proc. Lond. Math. Soc.} {\bf 4}, 381.

\smallskip\noindent
Clifford, W.K. (1878).	
{\it Am. J. Math.} {\bf 1}, 350.

\smallskip\noindent
Collins, R. E. (1992).
{\it Found. Phys. Lett.} {\bf 5}, 63.

\smallskip\noindent
Emch, G. (1963).
{\it Helv. Phys. Acta} {\bf 36}, 739, 770.

\smallskip\noindent
Finkelstein, D. {\it et al.} (1962).
{\it J. Math. Phys} {\bf 3}, 207.

\smallskip\noindent
Frobenius, F.G. (1878).
{\it Jo. Reine Angew. Mat.} {\bf 84}, 1.

\smallskip\noindent
Heslot, A. (1985).
{\it Phys. Rev.} D {\bf 31}, 1341.

\smallskip\noindent
Hucks, J. (1993).
{\it J. Math. Phys.} {\bf 34}, 5986.

\smallskip\noindent
Jammer, M. (1974).
{\it The Philosophy of Quantum Mechanics},
Wiley, New York.


\smallskip\noindent
Jauch, J. (1973).
In {\it The Physicist's Conception of Nature}, J. Mehra, ed., Reidel,
Dodrecht, pp. 300--319.

\smallskip\noindent
Jones, K.R. (1992).
{\it Phys. Rev.} D {\bf 45}, 2590.

\smallskip\noindent
Kunstatter, G. {\it et al.} (1983).
{\it J. Math. Phys.} {\bf 24}, 886.

\smallskip\noindent
Millard, A.C. (1997).
{\it J. Math. Phys.} {\bf 38}, 6230.

\smallskip\noindent
Nagasawa, M. (1993).
{\it Schr\"odinger Equations and Diffusion Theory},
Birkhauser, Boston.

\smallskip\noindent
Nash, C.G., and G.C. Joshi (1992).
						 {\it Int. J. Theor. Phys.} {\bf 31}, 965.

\smallskip\noindent
Nelson, E. (1966).
{\it Phys. Rev.} {\bf 150}, 1076.

\smallskip\noindent
Pauli, W. (1933).
{\it Z. Physik} {\bf 80} 573.

\smallskip\noindent
Pelce, P. (1996).
{\it Eur. J. Phys.} {\bf 17}, 116.

\smallskip\noindent
Porteous, I. (1981).
{\it Topological Geometry}, 2nd ed.,
Cambridge University Press.

\smallskip\noindent
Rowe, D.J., A. Ryman, and G. Rosensteel (1980).
             {\it Phys. Rev.} A {\bf 40}, 2362.

\smallskip\noindent
Strochi, F. (1966).
{\it Rev. Mod. Phys.} {\bf 38}, 36.

\smallskip\noindent
Stueckelberg, E.C.G.,  {\it et al.} (1960).
{\it Helv. Phys. Acta} {\bf 33}, 727.

\smallskip\noindent
Stueckelberg, E.C.G., and M. Guenin (1961).
{\it Helv. Phys. Acta} {\bf 34}, 621.

%

\end{document}